\documentclass[12pt]{article}
\usepackage{geometry}
\usepackage{amsfonts}
\usepackage{amsmath}
\usepackage{amssymb}
\usepackage{amsthm}
\usepackage{mathrsfs}
\usepackage{bbm}
\usepackage{bm}
\usepackage{cancel}
\usepackage{graphicx}
\usepackage[usenames,dvipsnames]{color}
\usepackage{fancyhdr}
\usepackage{mathrsfs}
\usepackage{wrapfig}

\def\q{\sf{Q$^N$}~}
\def\qn{\sf{Q$^N$}}
\def\pshs{\sf{PS}~}
\def\pshsn{\sf{PS}}
\def\shs{\sf{S}~}
\def\shsn{\sf{S}}
\def\Wss{\sf{W}~}
\def\Wssn{\sf{W}}

\def\Cssn{\sf{C}}
\def\shspv{ $\sf{S} \times \mathbbm R^+$}

\def\pshspv{ $\sf{PS} \times \mathbbm R^+$}

\def\Eucl{\sf{Eucl}~}

\def\Sim{\sf{Sim}~}
\def\Simn{\sf{Sim}}

\def\Riemn{\sf{Riem}}
\def\Diff{\sf{Diff}~}

\def\Conf{\sf{Conf}~}

\def\Dil{\sf{Dil}~}

\def\Transln{\sf{Transl}}

\def\nb{$N$-body problem~}

\def\bq{\begin{equation}}
\def\ee{\end{equation}}

\def\e{\emph}

\newcommand{\st}[1]{\textrm{\tiny{#1}}}

\def\03{\cite{Barbour2003~}}
\def\03n{\cite{Barbour2003}}
\def\82{\cite{Barbour1982}}
\def\82n{\cite{Barbour1982n}}
\def\rm{\textrm}
\def\bf{\textbf}
\def\sf{\textsf}

\def\3d{three-dimensional~}
\def\4d{four-dimensional~}

\newcommand{\Rn}[1]{$\mathbbm R^{#1}$}

\usepackage{hyperref}

\hypersetup{
   colorlinks=true,         % false: boxed links; true: colored links, false is default
   linkcolor=blue,          % color of internal links, red is default
   citecolor=red,        % color of links to bibliography
   urlcolor=Violet            % color of external links, cyan is default
}

\begin{document}

\title{ \large \bf{The Solution to the Problem of Time in Shape Dynamics} }

%\title{\Large  Shape Dynamics solves the Problem of Time}

%\author{\bf{ Julian Barbour$^{1,2}$
%{\rm,}
%Tim~Koslowski$^{3}$
%{\rm and}
%Flavio~Mercati$^{4}$}}
%%\email{barbourj@physics.ox.ac.uk}
%%\email{t.a.koslowski@gmail.com}
%%\email{fmercati@perimeterinstitute.ca.}
%\affiliation{\it $^1$College Farm, South Newington, Banbury, Oxon, OX15 4JG, UK,\\
%$^2$Visiting Professor in Physics at the University of Oxford,\\
%$^3$ University of New Brunswick, Fredericton, NB, E3B 5A3, Canada,\\
%$^4$Perimeter Institute for Theoretical Physics, 31 Caroline Street North  Waterloo, ON N2L 2Y5, Canada.}

\author{{Julian Barbour,$^{1,2}$
Tim~Koslowski$^{3}$
{\rm and}
Flavio~Mercati$^{4}$}\vspace{12pt} \\
%\email{barbourj@physics.ox.ac.uk}
%\email{t.a.koslowski@gmail.com}
%\email{fmercati@perimeterinstitute.ca.}
\it \small $^1$College Farm, South Newington, Banbury, Oxon, OX15 4JG UK,\\
\it \small $^2$Visiting Professor in Physics at the University of Oxford, UK\\
\it \small  $^3$ University of New Brunswick, Fredericton, NB, E3B 5A3 Canada,\\
\it \small $^4$Perimeter Institute for Theoretical Physics, 31 Caroline Street North,\\
\it \small Waterloo, ON, N2L 2Y5 Canada.}
\date{}

\maketitle

\begin{abstract}
%The absence of unique time evolution in General Relativity (GR) in its original spacetime formulation leads to the as yet unsolved `problem of time' in quantum gravity. We show that in Shape Dynamics (SD), an observationally equivalent dual representation of Einstein's theory, there is unique evolution with respect to a monotonic dimensionless independent variable that serves as `time'. This is achieved through the observation that all objective predictions of GR for a spatially closed universe are indistinguishable from predictions derived from a dimensionless and scale invariant theory that evolves wrt a preferred independent variable. This equivalence solves the classical problem of time and is a promising basis for attacking the quantum problem. 

%The absence of unique time evolution in Einstein's spacetime description of gravity leads to the hitherto unresolved `problem of time' in quantum gravity. Shape Dynamics is a representation of gravity that is objectively equivalent to General Relativity but trades spacetime refoliation invariance for three-dimensional conformal invariance. We present here the logical completion of Shape Dynamics, which gives a \emph{dimensionless} description of gravitational dynamics. We show that in this framework the classical problem of time is completely solved. Since this is impossible within the spacetime description, we believe Shape Dynamics provides a promising basis for solution of the quantum problem and the creation of quantum gravity.

The absence of unique time evolution in Einstein's spacetime description of gravity leads to the hitherto unresolved `problem of time' in quantum gravity. Shape Dynamics is an objectively equivalent representation of gravity that trades spacetime refoliation invariance for three-dimensional conformal invariance. Its logical completion presented here gives a \emph{dimensionless} description of gravitational dynamics. We show that in this framework the classical problem of time is completely solved. Since a comparable definitive solution is impossible within the spacetime description, we believe Shape Dynamics provides a  key ingredient for the creation of quantum gravity.
\end{abstract} 

\maketitle

\section{Introduction}
%\bf{Introduction.} 
In the canonical formulation of General Relativity due to Arnowitt, Deser and Misner (ADM),
the total Hamiltonian is a pure constraint:
\bq
{\mathcal{H}}=\int \rm dt \int \rm d^3x \left( N(x,t) \,  H(x) -2 N_i(x,t)\, \nabla_j p^{ij} \right) \,,
\label{adm}
\ee
in the case of spatial slices that are compact without 
boundary.
Variation wrt the position-dependent lapse $N$ gives at each space point the constraint
\bq
 H = \frac 1 {\sqrt g } \left( p^{ij}p_{ij}-{1\over 2}p^2 \right) - \sqrt g \, R=0,\label{ham}
\ee
where $p^{ij}$ are the momenta conjugate to the 3-metric $g_{ij}$, $p=g_{ij}p^{ij}$, and $g= \det g_{ij}$.
Variation wrt the shift $N_i$ leads to the linear ADM momentum constraint
\bq
-2 \, \nabla_j p^{ij}=0  \,.\label{ladm}
\ee
It generates 3-diffeomorphisms
and is a proper gauge constraint. In contrast, the quadratic constraint (\ref{ham}) \emph{evolves} the system but at arbitrarily different rates at each space point depending on the freely specifiable
lapse $N$ (many-fingered time). This lack of determinate evolution arises from the refoliation invariance of General Relativity (GR) in the spacetime representation and for quantization creates the severe and long unresolved \emph{problem of time} \cite{Anderson:review_pot}. It has two core aspects. 

%
%In quantum mechanics (QM) time is not an observable, it's an
%independent variable with respect to which the state of the system
%evolves. Time is \e{sui generis}. 
%Conceptually, GR presents problems for quantization
% because Einstein made time dynamical and did away with simultaneity.
%The problem of time has several  aspects \cite{Anderson:review_pot}, but
%in this letter we want to concentrate on what we identify as the
%core of the problem, which is  twofold:

%1. The quadratic constraint (\ref{ham}) is local%, and there
%%is a different constraint at each point in space $x$
%. The time parametrization can be freely chosen \emph{locally}, and the spatial slices can evolve at different 
%rates at each point. This changes the foliation of spacetime,
%1. Each choice of foliation corresponds to a different curve in phase space.
%One cannot treat the Hamiltonian constraint of GR as any gauge constraint,
%and quotient phase space to obtain the physical degrees of freedom.
%%To do so, one would have to solve the equations of motion.
%In the words of Dirac \cite{Dirac:CMC_fixing}:
%\begin{quote} 
%\emph{Thus we have the situation that we cannot specify the initial state for a 
%problem without solving the equations of motion. The formalism is thus not suitable 
%to dealing with practical problems.}
%\end{quote}
%This issue is often deemed ``problem of many-fingered time''.
%This problem is serious already at the classical level. It is 
%the reason why GR fails to satisfy the Poincar\'e principle (see below),
%in both its strong and weak form.

1. Unlike (\ref{ladm}), the Hamiltonian constraints (\ref{ham}) do not arise from the action of a group on configuration space, which is why GR fails to satisfy the Mach--Poincar\'e principle (see below)
in either its strong or weak form. Instead, as Dirac noted long ago \cite{Dirac:CMC_fixing}, the Hamiltonian constraints entangle gauge invariance with true dynamics. %Moreover, because it would violate closure of the constraint algebra no single smearing $H(N)$ can be taken as a physical Hamiltonian rather than a constraint.
%Thus, it is not a proper gauge constraint, and one cannot quotient phase space to obtain physical degrees of freedom. To do so, one would have to solve the equations of motion first  \cite{Dirac:CMC_fixing}.
%This is already serious at the classical level. 
Shape Dynamics (SD) disentangles them, without
introducing either second class constraints or additional degrees of
freedom, by replacing the Hamiltonian constraints with local three-dimensional conformal constraints and a single volume constraint that acts as the generator of dynamics. This eliminates many-fingered evolution.

2. Even in a particular foliation, a quantum problem remains. The theory is still reparametrization
invariant with a residual global constraint corresponding to a vanishing Hamiltonian.
This feature is not specific to GR and arises in any theory with a quadratic constraint. Traditional quantization automatically leads to a timeless model with a time-independent Schr\"odinger
 equation. This is the quantum `frozen-formalism' problem of time. We solve it by finding a dimensionless and scale-invariant system with a preferred independent variable,\footnote{We say `independent variable' rather than `time' because, as explained in footnote 8, $\tau$ has no relation at all to the readings of physical clocks.} $\tau$, that makes predictions indistinguishable from the objective predictions of GR. The transition to a representation in which both the true dynamical degrees of freedom and, in our key innovation, $\tau$ are dimensionless is the logical completion of SD and simultaneously the solution to the problem of time.
 
 \section{Approaches to the Problem of Time}

To highlight the new elements we bring to the issue, we briefly review the most sophisticated current line of attack, the `effective approach' \cite{Bojowald:effective_pot}.
A timeless quantum universe is accepted as fundamental and assumed to be in a semiclassical high-quantum-numbers regime with its wavefunction $\Psi$ peaked around a classical trajectory. One physical observable is singled  out as an internal time wrt which $\Psi$ evolves. Since no observable possesses everywhere the `good-clock'
requirement of monotonicity, a `grasshopper' strategy is adopted: the internal time `jumps' from one variable to another on the approach to a turning point.

%The problem with physical internal clocks is that in QM time is \emph{not} observable: it is just a real parameter wrt which the wavefunction evolves, much like Newton's
%absolute time. The observables are linear operators with a rich quantum structure (variance, correlations, etc). We find the `unquantizing' of an observable by making it an internal clock
%questionable.

While the universe may well get into a semiclassical regime, peaking of $\Psi$ around a particular classical trajectory is a much stronger assumption. Will the required `unquantizing' of a degree of freedom occur?

A much simpler solution would be one in which, just as in existing quantum mechanics, time is \e{sui generis}, i.e., quite different from the quantum degrees of freedom because it is a genuine independent variable. We shall show that such a variable does exist in a spatially closed universe and moreover is monotonic.

The demonstration relies on two facts. First, the objectively equivalent shape-dynamic representation of Einstein's theory shows that there is a mathematically and physically well-defined notion of simultaneity within closed-space GR. This solves the foliation problem. Second, the shape-dynamic variables divide unambiguously into two kinds: dimensionless true degrees of freedom and the \e{single} monotonic dimensionless independent variable $\tau$.

%
%
%In this letter we completely get round the problem by identifying an `internal' time in GR which
%is not a Dirac observable. This allows us the luxury to formulate some very restrictive criteria 
%for a satisfactory solution of the problem of time: We require the dynamics to be generated
%by a physical Hamiltonian, a time variable which  grows monotonically
%along classical trajectories and which \emph{is not a observable,} and finally no mixing-up of gauge and dynamical evolution: the Hamiltonian should
%be gauge-invariant and act on the reduced phase space only.
\section{Shape Dynamics}

Shape Dynamics has been developed in stages over more than 30 years \cite{barbourbertotti:mach,Barbour_Niall:first_cspv,barbour_el_al:physical_dof,gryb:shape_dyn}. It is a general framework for all closed dynamical systems in which there exists a reduced configuration space 
%\shs
$\sf S$ 
(Shape Space) of physical degrees of freedom that, together with the independent variable, are all \emph{dimensionless}. This is the logical completion referred to in the abstract.

Its foundation is the Mach--Poincar\'e principle
%(\MPprinciplen) 
\cite{poincare:principle,barbourbertotti:mach,Barbour:DefMach}: \emph{Initial data in} \shs \emph{in the form of a point and a direction (strong form) or a point and tangent vector (weak form) must determine the physical evolution curve in \shs uniquely.}

General Relativity fails to realize the Mach--Poincar\'e principle in Wheeler's superspace \Wssn$=$\Riemn/\Diff \cite{WheelerSuperspace67,Giulini:superspace}:
 given a
point and direction (or tangent vector) in \Wss
there is a whole \emph{sheaf} of solutions to the ADM equations
of motion, each corresponding to a different foliation of one and the same spacetime.
York's work on the initial-value problem in GR \cite{York:cotton_tensor,York:york_method_prl} 
hinted that \emph{conformal superspace} \Cssn, the space of conformal
3-geometries, is the physical configuration space of gravity, not \Wssn.
%All but one global mode of the quadratic constraints are
%traded for a new linear constraint
%\begin{equation}
%g_{ij} \, p^{ij} - \int d^3 x \, g_{ij} \, p^{ij} / \int d^3 x \, \sqrt g  = 0 \,,
%\end{equation}
%generating volume preserving conformal transformations
%
%%
%The definition of a variational principle on this configuration
%space, based on best-matching, 
%
This does not quite work: the implicit requirement of full conformal invariance freezes the dynamics. One is forced to a less strict notion of what is dynamical and to
allow the \emph{total volume} of a compact 3-geometry, $V = \int  d^3 x \, \sqrt g$, to evolve \cite{barbour_el_al:physical_dof,Barbour:new_cspv,gryb:shape_dyn}. Then the strong Mach--Poincar\'e principle
is satisfied on
conformal superspace \emph{plus volume}, with a single simultaneity-defining global Hamiltonian
constraint. The final step to the logical completion of SD, anticipated in \cite{Koslowski:ObservableEquivalence} and carried out explicitly in this paper, is to deparametrize the residual global constraint. This completes the solution to the problem of time.

We expect that many readers, convinced by Einstein and Minkowski that universal simultaneity cannot be meaningfully defined, will balk at our solution.  However, we emhasize that, despite its
ontologically preferred time, SD creates the familiar spacetime picture on-shell. It therefore reproduces all the confirmed predictions of GR and moreover gives a much simpler account of gravitational dynamics: one shape of the universe succeeds another along a unique evolution curve in $\sf S$.

\section{\label{bb} The $N$-Body Quantum Problem of Time}
%\bf{The $N$-Body Quantum Problem of Time.} 
We use  the Barbour--Bertotti (BB) formulation of the \nb  \cite{barbourbertotti:mach,JuliansReview} as a toy model  to illustrate our key innovation, the insistence that both the independent variable and the observables be \emph{dimensionless}.
The BB extended configuration space is $\mathbbm{R}^{3N}$; $\bf{r}_a = (r_a^1 ,r_a^2,r_a^3) \in \mathbbm{R}^{3}$ are the Cartesian coordinates of particle $a \in \{ 1,\dots,N\}$.
Newton's absolute space is eliminated by a reduction to \qn~$=$ \Rn{3N}/\Eucl, where \Eucl
is the Euclidean group of rigid translations and rotations \cite{barbourbertotti:mach}. 
Newton's absolute time is replaced by a path label $\lambda$ in a geodesic principle on \qn.
The resulting theory is invariant under $\lambda$-dependent \Eucl transformations \cite{barbourbertotti:mach,JuliansReview} and reproduces Newtonian gravity but
with the further predictions that
\begin{equation}
\bf P = \sum_{a=1}^N \bf p^a=0 \,, ~~~ \bf L  = \sum_{a=1}^N \bf r_a \times  \bf p^a=0\,,\label{mcon}
\end{equation}
where $\bf p^a$ are canonical momenta $\{ r_a^i  , p^b_j \} = {\delta_a}^b  \, {\delta^i}_j $, and $\bf P$ and $ \bf L$ are the total
linear and angular momenta of the system. The linear constraints (4) taken together are analogous to the ADM
momentum constraint (\ref{ladm}). The  geodesic principle on \q gives rise to a quadratic reparametrization constraint that fixes the magnitude of the canonical momenta $ \bf{p}^a$ and is the counterpart of ADM's (\ref{ham}):
\bq
H = \sum_{a=1}^N {\bf{p}^a\cdot\bf{p}^a\over 2 \, \mu_a}-E+V_{\st{New}}=0.\label{qcon}
\ee
Here  $\mu_a = m_a /M$, $ M = \sum_b m_b$, are dimensionless `geometrical' masses, and $V_{\st{New}}$ is the Newton potential:
\bq
V_{\st{New}}=- \sum_{a<b} {\mu_a \mu_b\over r_{ab}} \,, ~~~ r_{ab} = ||\bf{r}_b-\bf{r}_a|| \,. \label{npot}
\ee 
Note that $[V_{\st{New}}]=\ell^{-1}$ (there are no time or mass dimensions in our dynamical variables), $[\bf r_a] = \ell$ and (from (\ref{qcon})) $[\bf p^a] = \ell^{- \frac 1 2}$.

To implement the Mach--Poincar\'e principle, we postulate that only 
invariants of the similarity group \Sim (\Eucl plus dilatations) are true Lagrangian degrees of freedom (from which, we anticipate, quantum Dirac observables are to be constructed) and reduce our configuration space
to Shape Space \shsn~$=$ \Rn{3N}/\Simn. The scale-invariance constraint analogous to (\ref{mcon}) would be vanishing of the \e{dilatational momentum} $D$%($[D]=[\bf L]=\ell^{1/2}$):
\bq
D=\sum_{a=1}^N \bf r_a\cdot \bf p^a=0 \,.\label{dcon}
\ee
We say `would be' because, unlike the constraints (\ref{mcon}), $D=0$ is not conserved by the Newton potential (\ref{npot}), which is invariant under translations and rotations but not dilatations. In fact, $2\,D \approx \dot I_{\st{cm}}$ is the time derivative of the center-of-mass moment of inertia $I_{\st{cm}}$,
\bq
I_{\st{cm}}=\sum_{a=1}^N \mu_a ||  \bf r^\st{cm}_a ||^2 \equiv  \sum_{a<b} \mu_a \mu_b\,r_{ab}^2 \,, \label{cm}
\ee
where $\bf r^\st{cm}_a  = \bf r_a  - {\textstyle \sum_b} \mu_b \, \bf r_b$. Note that $R = I_{\st{cm}}^{1/2}$ measures the root-mean-square length, or scale, of the system ($[I_{\st{cm}}^{1/2}]=\ell$). Thus, whereas $\bf P$ and $\bf L$, which measure unobservable overall change of position and orientation in Newton's absolute space, can be made to vanish in BB, this cannot be done for the overall expansion rate $D$.

But total scale is purely conventional in a closed universe. Expansion of the universe is not seen but deduced -- from comparison of simultaneously observed wavelengths. One way to solve this problem is to change the potential in
such a way so as to ensure that the constraint (\ref{dcon}) is conserved. The resulting theory then satisfies the strong Mach--Poincar\'e principle and, as shown in \cite{BLMpaper}, has interesting quantum consequences, in particular, a breakdown of scale invariance due to a quantum anomaly. Here we take a different
route, presenting a theory which satisfies only the weak Mach--Poincar\'e principle but is still scale invariant.

% There are two ways to resolve the problem of scale -- and, as we shall see, time. We outline the more radical one first. 

%The putative Machian symmetry $D=0$ is broken because $H$ in (\ref{qcon}) is not scale covariant: the kinetic energy 
%$T =  \sum_{a=1}^N {\bf{p}^a\cdot\bf{p}^a\over {2 \mu_a}}$, $E$ and $V_{\st{New}}$ transform differently under $D$. An $[\ell^{-2}]$ potential would be needed, so we replace $E-V_{\st{New}}$ by $V_{\st{s}}= I^{-1/2}_{\st{cm}}V_{\st{New}}$ \cite{barbour:scale_inv_particles}. With it, all four Machian constraints hold: the quadratic (\ref{qcon}) and the linear $\bf{P}=0, \bf{L}=0, D=0$, and the constraint algebra closes.
%This theory has interesting quantum consequences \cite{BLMpaper}, in particular a breakdown of scale invariance
%due to a quantum anomaly. In \cite{BLMpaper} the possibility of linking this phenomenon with the emergence of time
%has been explored.

%Here we take the second route to scale invariance.

At any instant the Newtonian centre-of-mass kinetic energy $T$ decomposes uniquely into three terms \cite{SaariBook}: $T=T_{\st{r}}+T_{\st{d}}+T_{\st{s}}$, the rotational, dilatational and shape parts, respectively. We seek a theory with only $T_{\st{s}}$. The constraints  (\ref{mcon}) eliminate $T_{\st{r}}$ and the translational part $T_{\st{p}}$. In the next step, instead of changing the potential, we retain $E-V_{\st{New}}$ but swap the reparametrization invariance of (\ref{qcon}) for a time $\tau$ that subsumes $T_{\st{d}}$.

Our $\tau$ will be \e{monotonic} thanks to an important fact: if in any dynamical system the potential $V$ is homogeneous of some degree $k$, Euler's homogeneous-function theorem and Newton's second law imply $\ddot I_{\st{cm}}=4(E-V)-2kV$. Since $k=-1$ for $V_{\st{New}}$, this becomes $\ddot I_{\st{cm}}=4E-2V_{\st{New}}$. Now $V_{\st{New}}$ is negative, so if, as we shall assume, $E\ge 0$, then the function $I_{\st{cm}}(t)$ is concave upward and $D$ monotonic. This is not generically true of any physical observable. We already noted that scale, like time, is not observable; we now see that, through the monotonicity of $D$, scale shares another defining property of Newton's time $t$.

Homogeneity of the potential has another important consequence: dynamical similarity \cite{LandauLifshitz1}. Let a solution of the dynamical equations be given and the coordinates and time be scaled with a constant $C$ as follows: $\bf{r}_a \rightarrow C\, \bf r_a\,,t\rightarrow C^{1-k/2}$. Then geometrically similar paths are obtained. The elapsed times at which corresponding points are reached are in the ratio $t'/t=(l'/l)^{1-k/2}$. For the Newton potential with $k=-1$, this yields Kepler's third law.

Using the example of Newtonian dynamics, we can illustrate the way in which a given physical theory can be represented in two very different ways. Let solutions of Newton's equations be generated in $\mathbbm{R}^{3N}$ and then `projected' to Shape Space by abstraction of everything defined by conventional choice of units and the origin and orientation of the coordinate axes. Then the Newtonian solutions $\bf r_a(t)$ are mapped to a curve of shapes $s(\lambda)$ in \shsn. We call this the passage from the \emph{coordinatized} to the \emph{objective} representation. The latter eliminates all redundancy and human convention from the description of the dynamics. The description is \emph{dimensionless} and retains only what is objective: as we have seen, when the potential is homogeneous, a seemingly distinct one-parameter family of solutions in the coordinatized repesentation is mapped to a single objective curve in \shsn.

We now show how a suitably defined $\tau$ eliminates the dilatational part $T_{\st{d}}$. We should derive equations in \shsn, but, the non-Abelian rotations being difficult, we
%\footnote{To quotient wrt translations it's sufficient to go to the center-of-mass frame $\bf r_a^\st{cm}
%= \bf r_a - \sum_b \mu_b \bf r_b$, $\bf p^a_\st{cm} = \bf p^a - \frac 1 N \sum_b \bf p^b$, which reduces 
%the configuration space to $\mathbbm{R}^{3N-3}$.} 
quotient only wrt dilatations and translations, which takes us to `pre-shape space' \pshs=$\mathbbm{R}^{3N-3}$/\Dil$\times$\Transln, on which we introduce the `gauge-fixed' coordinates and momenta 
\bq
\bm \sigma_a=\sqrt{ \mu_a} \, {\bf r_a^\st{cm} \over R} \,, ~~~~  \bm \pi^a= { 1 \over \sqrt\mu_a }\, \frac{R}{D_0} \,\bf p^a_\st{cm}  -  \sqrt{\mu_a}\,\frac D {D_0} \,  \bm \sigma_a \,,~~~~R = I_{\st{cm}}^{1/2}.    \label{dm}
\ee
Here $\bm \sigma_a$ coordinatize a unit $(3N-4)$-dimensional sphere, and $\bm \pi^a$ are the $3N-4$ momenta tangent to that sphere divided by the value $D_0$ of $D$ at the point chosen to begin evolution.\footnote{In the many equivalent coordinatized representations, $D_0$ will have different values, but since we always take dimensionless ratios the same quantities are invariably obtained in the objective description. It is important that the point in $\mathbbm{R}^{3N}$ at which the evolution is begun, where $D=D_0$, corresponds to one and the same shape $s_0\in \sf{S}$.} This division by $D_0$ makes the momenta \e{dimensionless} like $\bm \sigma_a$. We have  
\bq
\begin{array}{cc}
\sum_\st{a=1}^\st{N}  \bm \sigma_a \cdot \bm \sigma_a = 1 \,, &  \sum_\st{a=1}^\st{N}  \bm \pi^a \cdot \bm \sigma_a = 0 \,,
\vspace{6pt}\\
\sum_\st{a=1}^\st{N}  \sqrt{\mu_a} \,  \bm \sigma_a = 0 \,, &  \sum_\st{a=1}^\st{N} \sqrt{\mu_a} \,  \bm \pi^a = 0 \,.  
\end{array}
\label{ConstraintsOnSigmaPi}
\ee
%Note especially the second equation in the first line and its near identity to the constraint (\ref{dcon}). It shows that in this treatment, in which the Newton potential is retained, we have also achieved scale invariance. Note also that the relations (\ref{ConstraintsOnSigmaPi}) hold in pre-shape space, while (\ref{dcon}) holds in the cotangent bundle of $\mathbbm{R}^{3N}$. 

The constraints  (\ref{ConstraintsOnSigmaPi}) are first class among themselves wrt the Hamiltonian constraint (\ref{qcon}).

We now express the Newton potential as $V_{\st{New}}(\bf r) = V_{\st{s}}(\bm \sigma) / R $, calling $V_{\st{s}}$ the \emph{shape} potential; it is a function on Shape Space.\footnote{As pointed out in \cite{BLMpaper}, the dimensionless $V_{\st{s}}$ is not only the potential that governs the scale-invariant dynamics but also an objective measure of gravitational complexity; we anticipate that it will be the most important Dirac observable in the quantum treatment.}  The kinetic energy decomposes into the dilatational part, $T_\st{d} = \frac 1 2 D^2/R^2$, and shape part
$T_\st{s} = \frac 1 2 D_0^2 \, K_\st{s}  / R^2$, where  $K_\st{s} =\sum_{a=1}^N \bm \pi^a\cdot \bm \pi^a $.
 The $(\bm \sigma_a, \bm \pi^b) $ coordinates are scale invariant, as they Poisson commute with $D$ and $R$,
\bq \label{ShapeVariablesInvariance}
\{ f(D,R) , \bm \pi^a \} = \{ f(D,R) ,  \bm \sigma_a \} = 0 \,,
\ee

These relations show that the associated Dirac bracket is block diagonal, which is important for the deparametrization performed below.

The  symplectic structure on pre-shape space is  
\bq
\begin{array}{l}
\{ \sigma^a_i ,  \pi_b^j \}  = D_0^{-1}  \left(  {\delta^a}_b \, {\delta^j}_i -  \sigma_a^i \, \sigma_b^j \right) \,,
 \\
\{\pi^a_i ,  \pi^b_j \}
=  D_0^{-1} \left( \sigma_a^i \, \pi^b_j -\sigma_b^j \,\pi ^a_i \right)  \,,  \label{PreShapeSpaceSymplStructure}
\\
\{ \sigma^a_i ,  \sigma_b^j \}  = 0 \,. 
\end{array}
\ee
%Some dimensional analysis: we have $[D] =\ell^{\frac 1 2}$, $[R] =\ell$; $[\bm\pi^a] =\ell^{\frac 1 2} $, $\bm \sigma_a$  and $V_{\st{s}}$  are dimensionless.

We now introduce the \e{monotonic and dimensionless independent variable} $\tau = D/D_0$ and obtain an unconstrained  (true) Hamiltonian. The $\log$ of $R$ is canonically conjugate to $D$, $\{ D , \log R \} = 1$.
%\footnote{One should write, for dimensional consistence, $ \log R/R_0$ where $R_0$ is a reference value %of $R$. But $ \log R/R_0 = \log R/R_1 + \log R_1 /R_0 $ and since $R_0$, $R_1$ Poisson commute with %everything, they drop out of the equations of motion.} 
Therefore $\mathcal H = - \log (R/R_0)+const $ can be used as the generator of $\tau$-translations:
% The $1 / D$ prefactor on the rhs of the canonical relations
% (\ref{PreShapeSpaceSymplStructure}) suggests a preferred choice for a time variable: 
% a `logarithmic' time $\tau = \log D/D_0$.  
take a shape observable $f= f(\bm \sigma , \bm \pi)$, its evolution wrt $D$ is
 \bq
\frac{\partial f}{ \partial \tau} = D_0 \, \{ \mathcal H  , f \} \,.
 \ee
Note that the $D_0$ on the rhs will cancel the $D_0^{-1}$ in (\ref{PreShapeSpaceSymplStructure}),
removing from the equations of motion all dependence on $D_0$, the only dimensionful 
quantity remaining in the theory. As is required for the objective description of a closed universe, \e{the equations of motion are dimensionless}.

The innovation of introducing a dimensionless time is, 
in our view, essential for the satisfactory passage from a constrained to a physical Hamiltonian. We are not aware that it has been considered hitherto. It leads to a notion of time very different from that of either Newton or Einstein. The origin of $\tau$  is conventional -- it can be taken at any point along the orbit in Shape Space. After that $\tau$ changes at a definite rate along the orbit from the origin as $R/R_0$ changes.\footnote{We should point out that any definite monotonic function of $D/D_0$ can also be chosen as time. We shall return to this point in footnote 8.\label{cav1}}  Besides monotonicity, the advantage of using $D$ to define $\tau$ is its status as a uniquely distinguished  \emph{collective} variable associated with the overall behaviour of the system. The growth of $\tau$ is created by all the physical degrees of freedom working together. This enables us to avoid arbitrary choice of individual physical degrees of freedom as `time' with the inevitable `hopping' from one to another.

From the conceptual point of view, it is also important that, interpreted in the usual scale-dependent coordinatized description, our variable $\tau$ appears as an internal time, but  it is an external evolution parameter in the scale-invariant dimensionless description. Since it is the latter which represents the objective state of affairs, the distinction, which appears in exactly the same way in geometrodynamic SD and will be a central feature in the quantum treatment, needs to be made.
 
To find the physical Hamiltonian $\mathcal H$ we solve the quadratic constraint (\ref{qcon}) for $R$:
\begin{equation} \label{EqForR}
R^2 \, E -  R \,   V_{\text{s}} -  \frac 1 2 \,  D^2 - \frac 1 2 \, D_0^2 \,K_\st{s}  = 0  \,.
\end{equation} 
In the simplest case when $E=0$ the Hamiltonian is
\begin{equation} \label{WMPH}
\mathcal H =  \log \frac{K_\st{s}  + \tau^2 }{| V_{\text{s}}|} \,,
\end{equation}
and the equations of motion it generates are (notice the disappearance of $D_0$): 
\bq \label{shapeEOMzeroenergy}
 \frac{d  {\bm \sigma}_a} {d \tau}   \approx   \frac{2 \,  {\bm \pi}^a}{{ K_\st{s}  + \tau^2 } }   \,, ~ \frac{d  {\bm \pi}^a} {d  \tau}  \approx   \frac{\partial \log | V_{\st{s}}|}{\partial \bm \sigma_a}  - \frac{K_\st{s}  }{K_\st{s}   + \tau^2 } \bm  \sigma_a \,.
\ee
Quotienting wrt rotations would not change anything essential, as
all our equations are rotationally invariant.

%Note that the $\tau$ dependence in (\ref{shapeEOMzeroenergy}) is not in any way analogous to explicit time dependence in laboratory physics induced by, say, a time-dependent magnetic field. What counts is that all allowed dynamical solutions in \shs are generated intrinsically in exactly the same way. Any point in the shape phase space (the cotangent bundle of \shsn) can be chosen as the intial point of evolution, at which by definition the value of $\tau$ is 1 (because $\tau=D/D_0=1$ at the initial point). At this point,  one specifies $6N-14$ rotationally 
%invariant components of $\bm \sigma_a (1)$ and $\bm \pi_a (1)$. The equations of motion (\ref{shapeEOMzeroenergy}) then generate the dynamical trajectories. One could also take $\tau\ne 1$ and still generate all the same trajectories. This is quite unlike the effect of a time-dependent external field in a laboratory, which will change the trajectories of the vector field in phase space.

The system of equations  on Shape Space that we have obtained, (\ref{shapeEOMzeroenergy}), is not autonomous. But this is not in any way analogous to explicit time dependence in laboratory physics
induced by, say, a time-dependent magnetic field. What counts are the complete dynamical
orbits in  \shsn. They represent the physical reality. To generate all the orbits that pass through a given point in $\sf S$, one can always adopt the convention that $\tau=1$ in the formulation of the initial-value problem. In it,
%, and two initial
%conditions that lie on the same orbit have to be considered equivalent. Therefore the
%initial value of $\tau$ does not enter the initial-value problem, because one can always adopt the convention that $\tau=1$ at the point where the
%initial-value problem is posed.  
one specifies
$6N-14$ rotationally  invariant components of $\bm \sigma_a (1)$ and $\bm \pi_a (1)$;  each 
choice of them will generate a different curve in \shsn. Taken together, all these  initial conditions 
generate all of the dynamical curves that pass through the given point. Nothing can change them. This is quite unlike what happens to dynamical evolution subject to time-dependent external fields.

Note also that in Newtonian terms the $\bm \pi_a (1)$ are the dimensionless ratios of the shape-changing momenta divided by the momentum $D_0$ in overall initial expansion.  

The theory defined by the Hamiltonian (\ref{WMPH}) satisfies the weak Mach--Poincar\'e principle because, given $\bm \sigma_a (1)$ and $\bm \pi_a (1)$, the equations of motion (\ref{shapeEOMzeroenergy}) generate a unique curve in \shsn. We can construct the Newtonian trajectory in the extended phase space $(\bf r_a , \bf p^a)$
by inverting (\ref{dm}) to obtain the scaled data $D(\tau)$ and $R(\tau)$ in Eq. (\ref{EqForR}). 
We must also specify $D_0$, but this, our only dimensionful quantity (it gives length dimensions
to everything else), amounts to the conventional choice of a unit.\footnote{We have already restricted the choice of units in the Newtonian representation by setting the Newton constant $G=1$ and using `geometrical' masses.} 

If $E \neq 0$, we need an extra initial datum:
the dimensionless $\epsilon = E \, D_0^2$, which appears in the Hamiltonian:
\begin{equation}\label{eno}
\mathcal  H =  \log \left[ V_{\st{s}}  + \sqrt{V_{\st{s}}^2 + {\textstyle \frac 1 2} \, \epsilon \, \left( K_\st{s}  + \tau^2 \right)} \right] \,.
\end{equation}
The Mach--Poincar\'e principle fails due to the necessity to specify $\epsilon \, \tau^2$, but the theory is still scale invariant and fully dimensionless with a monotonic independent variable. Our solution to the problem of time still holds.

%\e{Summary.} There exist \e{two} theories with all three linear gauge constraints expected under the assumption that translations and rotation are gauge and scale is unphysical convention. The first theory has the scale-invariant potential $V_{\st{s}}$ and momenta constrined by (\ref{qcon}); the second retains the Newton potential $V_{\st{New}}$ and has unconstrained momenta. We obtained the latter by swapping invariances: reparametrization for scale invariance and time. The objective equivalence of the two systems is the key fact underlying our solution to the problem of time. 

\section{\label{GSD} Geometrodynamic Shape Dynamics}
%\bf{Geometrodynamic Shape Dynamics.} 
From the point of view of the problem of time, the vitally important feature of geometrodynamic SD is that it completely eliminates many-fingered evolution and gives Einsteinian gravitational dynamics an architectonic structure essentially identical to $N$-body dynamics. As shown in \cite{gryb:shape_dyn}, it does this by trading all but one linear combination of the quadratic constraints at each space point in the integral (\ref{ham}) for a
\emph{volume-preserving conformal constraint} 
 \bq
 g_{ab} \, p^{ab} - \sqrt g \, Y = 0 \,,  \label{VPCTconst}
 \ee
 which together with the diffeomorphism constraints (\ref{ladm}) reduces the configuration space to
\shspv; \shs is conformal superspace \Cssn, the geometrodynamic shape space, and $ \mathbbm R^+$ represents the volume
 $V = \int d^3 x \sqrt g $, which is invariant under the transformations generated by (\ref{VPCTconst}) and, at this stage, a dynamical
 degree of freedom;
  $Y = \int d^3 x \, g_{ab} p^{ab} /V$ is proportional to `York time' and is monotonic along spacetime solutions that can be foliated by spacelike hypersurfaces of constant-mean-(extrinsic) curvature (CMC) \cite{York:york_method_prl}.\footnote{\label{co} In vacuum GR, the analogue of the passage from the coordinatized to the objective description of Newtonian dynamics that we described in Sec.~\ref{bb} amounts to identification of the conformal three-geometries on successive CMC slices of a spatially closed Einsteinian spacetime and projection of them to conformal superspace, where they form a unique curve that we regard as the entire objective content of theory.} The volume and $Y$ are canonical conjugates: $\{ V , Y \} = \frac 3 2$. 
  
Here too we find it convenient not to solve the linear (diffeo) constraint,
and we work in pre-shape space plus volume: \pshspv, where \pshs = \Riemn/\Conf.

The symmetry trading procedure allows one to trade the constraints
(\ref{ham}) for (\ref{VPCTconst}) with the \emph{exception of one single constraint.}
This eliminates the many-fingered evolution, the residual constraint becoming a reparametrization constraint for dynamics on \pshspv.
Exactly as before, we now describe the dynamics on \pshsn, 
using $Y$, like $\tau$, as independent variable and $V$ as physical Hamiltonian. Solved for $V$, the residual constraint gives
\bq
\mathcal H [g_{ab}, p^{ab} , Y] = \int d^3 x \, \sqrt g \, \phi^6[g_{ab},p^{ab} ,Y;x) \,, \label{SDhamiltonian}
\ee
where $\phi$ solves the Lichnerowicz--York (LY) equation (which always has
a unique solution \cite{Niall_73})
\bq
%2 \, \Lambda \, V^2 
\frac { g_{ac} \, g_{bd} \, p^{ab}_\st{T} \, p^{cd}_\st{T} } {\phi^{12} \,  g } -   \phi^{-4}  \left(   R - 8 \,  \phi^{-1}   \nabla^2   \phi \right)  - \frac 1 6 \, Y^2 = 0 \,, \label{LYequation}
\ee
$p^{ab}_\st{T} = p^{ab} - \frac 1 3 \, g^{ab} \, p^{cd} \, g_{cd}$ is the traceless part of $p^{ab}$.

The Hamiltonian (\ref{SDhamiltonian}) is obviously diffeo-invariant and also \e{fully} conformally invariant if $Y$ is treated as the independent variable,
\bq \label{ConfInv}
\mathcal H [\omega^4 g_{ab},  \omega^{-4}  p^{ab} , Y] =\mathcal H [g_{ab}, p^{ab} , Y] \,,
\ee
because (\ref{LYequation}) shows that $\phi[g_{ab},p^{ab} ,Y;x) $ transforms as
\bq \label{ConfTransfLYeq}
\phi[\omega^4 g_{ab},\omega^{-4}  p^{ab} ,Y;x)  = \omega^{-1}(x) \, \phi[g_{ab},p^{ab} ,Y;x) \,,
\ee
while the $\sqrt g$ factor in  (\ref{SDhamiltonian})  transforms as $\sqrt g \to \omega^6 \, \sqrt g$,
cancelling $\omega$.  Further, as noted in \cite{Barbour:new_cspv}, the LY equation (\ref{LYequation}), like the $N$-body equations,  exhibits a form of \emph{dynamical similarity}, which we here express in the more convenient equivalent form
%\footnote{In \cite{Barbour:new_cspv} eq. (\ref{DynamicalSim}) has been written in the alternative way %$\phi[\alpha^2 g_{ab},  p^{ab} ,\alpha^{-1} Y;x)  = \phi[g_{ab},p^{ab} ,Y;x) $,
%but we find the other way more useful.}
\bq \label{DynamicalSim}
\phi[ g_{ab}, \alpha^4 p^{ab} ,\alpha^{-2} Y;x)  = \alpha \, \phi[g_{ab},p^{ab} ,Y;x) \,,
\ee
where $\alpha$ is a spatial constant. This means that the
SD Hamiltonian scales covariantly with $\alpha^6$:
\bq
\mathcal H [g_{ab}, \alpha^4  p^{ab} , \alpha^{-2} Y] = \alpha^6 \, \mathcal H [g_{ab}, p^{ab} , Y] \,.
\ee
Note that the ADM Hamiltonian (\ref{ham}) \e{has an indefinite kinetic
energy} due to the $- \frac 1 2 p^2 $ term. Our description \e{completely removes} this long-standing conformal-factor problem. The volume-preserving conformal constraint (\ref{VPCTconst}) makes most of the position-dependent $- \frac 1 2 p^2 $ term pure gauge.
The remaining part, the spatial average of $p$, is used as the independent variable and therefore
removed from the dynamics. We are left with an SD Hamiltonian with non-negative
kinetic energy.

For the dimensional analysis, we use Dicke's convention \cite{DickeDimensions}, in which the 3-metric has $[g_{ij}] = \ell^2$,
$[g^{ij}] = \ell^{-2}$, while the coordinates (and, accordingly, space derivatives)
are dimensionless labels for points. The momenta are dimensionless $[p^{ij}] = 1$, the York time has 
$[Y] = \ell^{-1}$. 

We want to express the dynamics in terms of \emph{dimensionless} and conformally invariant
variables on the phase space of pre-shape space \pshsn. By analogy with the particle model, we choose the unimodular metric $\tilde g_{ab} =  g_{ab} / g^{\frac 1 3}$ (analogous to $\bm \sigma_a$) and dimensionless shape momenta $\tilde p^{ab}_\st{T} = Y_0^2 \, g^{\frac 1 3} \, p^{ab}_\st{T}$, which have exactly the same structure as $\bm p^a$: they are obtained by multiplying
the traceless part of $p^{ab}$ by $g^{\frac 1 3}$ (the analogue of $R$) to make it conformally invariant, and
then multiplying by the appropriate power of $Y_0$, an initial value of $Y$ (analogous to $D_0$), to make it dimensionless.
In analogy with Eq. (\ref{ShapeVariablesInvariance}), the shape variables $(\tilde g_{ab} , \tilde p^{ab})$ Poisson commute with every functional of $Y$ and $V$,
\bq
\{ F[Y,V] , \tilde g_{ab} \} = \{ F[Y,V] , \tilde p^{ab} \} = 0\,,
\ee
and satisfy constraints analogous to (\ref{ConstraintsOnSigmaPi}):
\bq
\det \tilde g_{ab} = 1 \,, \qquad  \tilde g_{ab} \, \tilde p^{ab} = 0\,.
\ee
The symplectic structure on \shs perfectly matches that of the particle model (\ref{PreShapeSpaceSymplStructure}),
\begin{align} \label{GeometrodynamicsSymplStructure}
\{ \tilde g_{ab} (x) , \tilde p^{cd}_\st{T}(y) \}   &= {\textstyle Y_0^2 \left( \frac 1 2 \delta^c_{(a} \delta^d_{b)} 
- \frac 1 3 \,  \tilde  g_{ab} \,\tilde   g^{cd} \right)  \delta(x,y) \,,}  \nonumber
\\
\{ \tilde p^{ab}_\st{T} (x) , \tilde p^{cd}_\st{T}(y) \}   &= {\textstyle \frac 1 3 Y_0^2 
\left( \tilde g^{cd} \, \tilde p^{ab}_\st{T} - \tilde g^{ab} \,  \tilde p^{cd}_\st{T}  \right) \delta(x,y) \,,}  \nonumber
\\
\{\tilde g_{ab} (x) , \tilde g_{cd} (y)  \} &=  0 \,.
\end{align}
The  SD Hamiltonian has the dimensions of a volume,  $[{ \mathcal H}] = \ell^3$, the Poisson brackets an inverse area $[\{ . , .\}]=\ell^{-2}$.

We now exploit the two invariances (\ref{ConfInv}) and (\ref{DynamicalSim}) to write a dimensionless Hamiltonian 
 $\tilde{\mathcal H}$  on \pshs that generates evolution in the dimensionless independent variable $\tau= Y/Y_0$:
\bq
\tilde {\mathcal H} [\tilde g_{ab}, \tilde p^{ab},\tau] = \int d^3 x \, \tilde \phi^6[\tilde g_{ab},\tilde p^{ab}_\st{T},\tau;x)\,,
\ee
where $\tilde \phi$ is a scalar \emph{density} of weight $\frac 1 6$ (it contains a factor of $g^{\frac 1 {12}}$) and solves the dimensionless equation 
\bq
\frac { \tilde g_{ac} \, \tilde g_{bd} \, \tilde p^{ab}_\st{T} \, \tilde p^{cd}_\st{T} } {\tilde  \phi^{12}} -  \tilde  \phi^{-4} \left(   \tilde R - 8 \, \tilde  \phi^{-1} \tilde   \nabla^2  \tilde  \phi \right)  - \frac 1 6 \, \tau^2 = 0 \,, 
\ee
where $\tilde R = R(\tilde g_{ab}) = g^{\frac 1 3} \, R(g_{ab})$, and $ \tilde   \nabla^2  =  \tilde g^{ab}  \nabla_a \nabla_b$.

A  functional $ F[\tilde g_{ab}, \tilde p^{ab} ] $ on \pshs then evolves as
\bq
\frac{d}{d \tau} F[\tilde g_{ab}, \tilde p^{ab} ] = \frac 2 3 \, Y_0^{-2} \{ \tilde {\mathcal H} [\tilde g_{ab}, \tilde p^{ab},\tau] , F [\tilde g_{ab}, \tilde p^{ab}]   \} \,,
\ee
and, as in the particle model, the $Y_0^{-2}$ term on the rhs compensates for the
corresponding factor in (\ref{GeometrodynamicsSymplStructure}).

As York showed \cite{York:york_method_long,Niall_73,York:york_method_prl}, the diffeomorphism
constraint decouples from the conformally invariant degrees of freedom and
can be solved independently. This happens because the equations of motion in \pshs are diffeo-invariant and, in turn, the diffeomorphism constraint is conformally covariant. This parallels the particle model, for which the dynamics in \shs is rotationally invariant and the angular-momentum constraint is scale invariant.

In the initial-value problem on \shsn, one specifies a unimodular 3-geometry 
($\tilde g_{ab}$ modulo diffeos) and unconstrained transverse-traceless momenta $\tilde p^{ab}_\st{TT}$. These are 4 degrees of freedom per point. The Hamiltonian $\tilde{\mathcal H}$ then generates a unique curve in \shsn, parametrized by $\tau$. 
This system satisfies the weak Mach--Poincar\'e principle. The ADM description is obtained by introducing the determinant 
$g = Y_0^{-6} \, \tilde \phi^{12}[\tilde g_{ab},\tilde p^{ab}_\st{T},\tau;x) $. Like $D_0$ in the particle model, the value given to $Y_0$ is conventional. All dimensions in the ADM representation arise from it.

\subsection*{Extensions} 

The bosonic matter of the standard model can be included in this picture
 by requiring the matter fields to transform with conformal weight zero \cite{MatterPaper}.
 The two key properties (\ref{ConfTransfLYeq}) and (\ref{DynamicalSim}) still hold, so the result goes 
 through unchanged. The only exception to this is
 a  cosmological constant $\Lambda$.  It resembles the $E\neq 0$ case
in the particle model. The LY equation gains a constant term in addition to $Y^2$, and (\ref{DynamicalSim})
no longer holds (this happens also for a Higgs potential): one must rescale $\Lambda$ and
specify the additional dimensionless quantity $\lambda = \Lambda / Y_0^2$, so the Mach--Poincar\'e principle fails. 

This failure in no way affects our solution to the problem of time. It is rather to be seen as a potential criterion for theory selection. Einstein was always struck by the fact that GR is the simplest non-trivial theory of dynamical Riemannian geometry. Faith in simplicity would lead one to seek to explain the apparent existence of a cosmological constant $\Lambda$ through some process by which it emerges from a fundamental theory that does not contain it. In this connection, our assumptions of positive energy in the $N$-body problem and positive $\Lambda$, under which $\tau$ is certainly monotonic, are mild and supported observationally -- a fundamental $\Lambda$, if it does exist, is positive.

\section{Conclusions}
We believe that, taken together, the results of this paper solve the long-standing classical problem of time and are a promising basis for the quantum treatment. In Einsteinian gravity, the solution consists of two stages: disentanglement of the evolution from gauge followed by deparametrization. In the $N$-body problem, only deparametrization is needed. Let us recapitulate the key aspects.

 1. Geometrodynamic SD as established in \cite{gryb:shape_dyn} cleanly disentangles the true dynamical evolution in gravity from the gauge invariance associated with the coordinatized spacetime description, reducing the problem of time to reparametrization invariance. The disentanglement is simultaneously the key step to the representation of Einsteinian gravity in dimensionless form, the \emph{sine qua non} for the subsequent introduction of $\tau$ and the full solution to the problem of time. That simply cannot be done without the passage from Wheeler's metric-based superspace, with its \emph{three} degrees of freedom per space point, to metric-free conformal superspace, which, as York \cite{York:york_method_prl} noted, is dimensionless and has \emph{two} degrees of freedom per space point.\footnote{We should place on record that our paper, like all previous work on SD, draws heavily on York's work, in which our collaborator \'O Murchadha played an important role. In fact, York came rather close to our solution to the problem of time. However, his analysis was based on full conformal transformations, not the volume-preserving transformations generated by (\ref{VPCTconst}). He also did not aim for a dimensionless `time', settling instead for the dimensionful York time. This is probably because he did not have the advantage of knowing the dynamical similarity (\ref{DynamicalSim}) of the LY equation, which was first recognized by \'O Murchadha \cite{Barbour:new_cspv} and led to our idea of introducing a dimensionless independent variable.}
 
 2. After this step, closed-space geometrodynamic SD, just like the Newtonian $N$-body problem, admits an objective description in completely dimensionless and scale-invariant terms. The evolution takes place wrt an unambiguously defined monotonic independent variable.\footnote{This statement is subject to the minor caveat noted in footnote \ref{cav1}: any monotonic function $f$ of $\tau$ will serve equally well. This explains why our independent variable is not related to the readings of physical clocks. The change from $\tau$ to any monotonic $f(\tau)$ multiplies the Hamiltonian by a corresponding lapse. The evolution curves in Shape Space are unchanged. If any $f(\tau)$ is in any way distinguished it is $T=\rm{log}\,(\tau)$, which allows one to set initial data at $T=0$ but perhaps more significantly makes the Hamiltonian $T$-independent in interesting asymptotic regimes. This might have technical advantages. But
fundamentally no specific choice of $f(\tau)$ has any ontological priority. We have to make \emph{some} choice and with it a corresponding Hamiltonian to express the law that generates the successive shapes in \shsn. The shapes alone are ontological.} The observational equivalence of what we have called the coordinatized and objective descriptions ensures that this solves the problem of time. 
 
We also want to recall the observation made after (24), which is that the ADM kinetic energy has a negative part. It has been a problematic issue for quantum-gravity research for decades, especially approaches based on path integrals, which as a result are unbounded below. This conformal mode problem has been considered in many approaches, see e.g. Ho{\v r}ava's \cite{Horava:lif_point} modified kinetic term. In our treatment, the troublesome contribution divides into a local part, which is pure gauge, and a spatial average which, made dimensionless, plays the role of the independent variable in our description and, as we have already noted, ceases to be dynamic. Thus, the kinetic-energy part of the physical Hamiltonian is shown to be positive definite. This appears to be a major bonus that comes with our solution to the problem of time.

To end, we should like to add some further arguments for the dimensionless representation of both Newtonian and Einsteinian gravity. 

Physicists are widely agreed that the results of laboratory experiments or theoretical calculations must, if they are to be objective, be expressed in dimensionless form. This requirement, whose extension to a dynamically closed universe can hardly be questioned, is the sole principle that underlies our identification of the independent variable $\tau$ once gauge has been disentangled from dynamics. 

Moreover, the elimination of redundancy and convention, especially external scale, reveals more than one might expect. One sees this already in Newtonian dynamics. Far from being the single theory it appears to be in the familiar representation, it is actually \emph{three}: one with energy $E=0$ and two more, with $E<0$ (which we have not considered) and $E>0$. These theories are governed by very different Hamiltonians (compare (\ref{WMPH}) and (\ref{eno})). Furthermore, a dynamically closed Newtonian universe cannot have different energies; it can just have one of those three different governing laws. There no such thing as a conserved
energy of a universe with different possible values determined by initial conditions. Finally, time -- as deeply rooted in intuition as the 19th-century aether -- is simply not there. The only `time' we can find is $D/D_0$.

If passage to the dimensionless description puts Newtonian `spacetime' in such a different light, we can surely expect the same to happen with Einsteinian spacetime. The way GR is normally taught is that extremalization of the Einstein--Hilbert action selects among all pseudo-Riemannian manifolds, each with a complete four-dimensional metric $g_{\mu\nu}$, those that are candidates for physical realization. 
However, we start with a vastly more restricted ontology: three-dimensional conformal geometries. We then postulate that a point and a tangent vector in conformal superspace, $\sf C$, must uniquely determine an evolution curve in $\sf C$.

This bare input is sufficient to `create' an entire Einsteinian spacetime endowed with local proper distance, local proper time and local inertial frames of reference. They are artefacts of the fundamental law that generates the curve of shapes in $\sf C$ and disappear when we return to the objective description as outlined in footnote~\ref{co}. We only think they are there because of the way the fundamental law makes one shape follow another, picking out a very special curve in $\sf C$ in doing so \cite{BarbourReductionistDoubts}. 

If progress requires us to give up relativity of simultaneity, viewing the status it has acquired as historically contingent, we think the price is worth paying. Let us remind the reader of Dirac's words in 1958 \cite{DiracHamiltonianDynamics}, who was so struck by the simplicity of the Hamiltonian treatment obtained by giving up the spacetime description that he commented: ``I am inclined to believe from this that four-dimensional symmetry is not a fundamental property of the physical world.'' We would slightly modify that and say the appearance of four-dimensional symmetry is a by-product of the way one shape follows another.

%1. One can still consider reparametrizations of time in the objective description of the system. It turns out that one parametrization is preferred in the sense that it leads to asymptotically time-independent Hamiltonian for gravity.
 
%2. The idea of having a completely scale invariant and dimensionless description of the objective dynamics of a system can be applied to the quantum mechanics of an isolated system.
 
%3. Our solution to the problem of time assumes that the shape dynamics description is in a sense more fundamental than the spacetime description of gravity. This has two consequences: (1) it allows for differences between shape dynamics and GR in certain circumstances. (2) The dynamical similarity (\ref{DynamicalSim}) that we used to show that the \MPprinciple is implemented breaks down if a scalar field potential (including a cosmological constant) is present. The failure of the system with a fundamental scalar field potential to satisfy the \MPprinciple could be seen as a hint that the fundamental description of the universe should not have such. Both ideas are currently investigated.

\vspace{12pt}
\bf{Acknowledgements.} TK was supported in part through NSERC.
JB was supported by  a grant from the Foundational Questions Institute (FQXi) Fund, a donor advised fund of the Silicon Valley Community Foundation on the basis of proposal FQXi Time and Foundations 2010 to the Foundational Questions Institute.
FM was supported by a grant from the John Templeton Foundation. The opinions expressed in this publication are those
of the authors and do not necessarily reflect the views of the John Templeton Foundation.
Research at Perimeter Institute is supported by the Government of Canada through Industry Canada and by the 
Province of Ontario through the Ministry of Economic Development and Innovation. We thank Boris Barbour for helpful comments on the text.


\begin{thebibliography}{99}

\bibitem{Anderson:review_pot}
E.~Anderson, ``{The problem of time in quantum gravity},''
  \href{http://arxiv.org/abs/1009.2157}{{\ttfamily arXiv:1009.2157 [gr-qc]}}.

\bibitem{Dirac:CMC_fixing}
P.~A.~M. Dirac, ``{Fixation of coordinates in the Hamiltonian theory of
  gravitation},'' \href{http://dx.doi.org/10.1103/PhysRev.114.924}{{\em Phys.
  Rev.} {\bfseries 114} (1959) 924--930}.

\bibitem{Bojowald:effective_pot}
M.~Bojowald, P.~A. H{\"o}hn, and A.~Tsobanjan, ``{An effective approach to the
  problem of time},''
  \href{http://dx.doi.org/10.1088/0264-9381/28/3/035006}{{\em
  Class.Quant.Grav.} {\bfseries 28} (2011) 035006},
  \href{http://arxiv.org/abs/1009.5953}{{\ttfamily arXiv:1009.5953 [gr-qc]}}.

\bibitem{barbourbertotti:mach}
J.~Barbour and B.~Bertotti, ``{Mach's principle and the structure of dynamical
  theories},'' {\em Proc. R. Soc. A} {\bfseries 382} no.~1783, (1982) 295--306.

\bibitem{Barbour_Niall:first_cspv}
J.~Barbour and N.~O'Murchadha, ``{Classical and quantum gravity on conformal
  superspace},'' \href{http://arxiv.org/abs/gr-qc/9911071}{{\ttfamily
  arXiv:gr-qc/9911071}}.

\bibitem{barbour_el_al:physical_dof}
E.~Anderson, J.~Barbour, B.~Z. Foster, B.~Kelleher, and N.~O'Murchadha, ``{The
  physical gravitational degrees of freedom},''
  \href{http://dx.doi.org/10.1088/0264-9381/22/9/020}{{\em Class. Quant. Grav.}
  {\bfseries 22} (2005) 1795--1802},
  \href{http://arxiv.org/abs/gr-qc/0407104}{{\ttfamily arXiv:gr-qc/0407104}}.

\bibitem{gryb:shape_dyn}
H.~Gomes, S.~Gryb, and T.~Koslowski, ``{Einstein gravity as a 3D conformally
  invariant theory},''
  \href{http://dx.doi.org/10.1088/0264-9381/28/4/045005}{{\em Class. Quant.
  Grav.} {\bfseries 28} (2011) 045005},
  \href{http://arxiv.org/abs/1010.2481}{{\ttfamily arXiv:1010.2481 [gr-qc]}}.

\bibitem{poincare:principle}
H.~Poincar{\'e}, {\em Science et Hypoth{\`e}se}.
\newblock Paris, 1902.

\bibitem{Barbour:DefMach}
J.~Barbour, ``{The definition of Mach's principle},''
  \href{http://dx.doi.org/10.1007/s10701-010-9490-7}{{\em Found. Phys.}
  {\bfseries 40} (2010) 1263--1284},
  \href{http://arxiv.org/abs/1007.3368}{{\ttfamily arXiv:1007.3368 [gr-qc]}}.

\bibitem{WheelerSuperspace67}
J.~A. Wheeler, ``Superspace and the nature of quantum geometrodynamics,'' {\em
  Battelle Rencontres} (1967) 242--307.

\bibitem{Giulini:superspace}
D.~Giulini, ``{The superspace of geometrodynamics},''
  \href{http://dx.doi.org/10.1007/s10714-009-0771-4}{{\em Gen. Rel. Grav.}
  {\bfseries 41} (2009) 785--815},
  \href{http://arxiv.org/abs/0902.3923}{{\ttfamily arXiv:0902.3923 [gr-qc]}}.

\bibitem{York:cotton_tensor}
James W.~York, Jr., ``{Gravitational degrees of freedom and the initial-value
  problem},'' \href{http://dx.doi.org/10.1103/PhysRevLett.26.1656}{{\em Phys.
  Rev. Lett.} {\bfseries 26} (1971) 1656--1658}.

\bibitem{York:york_method_prl}
J.~James W.~York, Jr., ``{Role of conformal three geometry in the dynamics of
  gravitation},'' \href{http://dx.doi.org/10.1103/PhysRevLett.28.1082}{{\em
  Phys. Rev. Lett.} {\bfseries 28} (1972) 1082--1085}.

\bibitem{Barbour:new_cspv}
J.~Barbour and N.~O'Murchadha, ``{Conformal Superspace: the configuration space
  of general relativity},'' \href{http://arxiv.org/abs/1009.3559}{{\ttfamily
  arXiv:1009.3559 [gr-qc]}}.

\bibitem{Koslowski:ObservableEquivalence}
T.~Koslowski, ``Observable equivalence between general relativity and shape
  dynamics,'' \href{http://arxiv.org/abs/1203.6688}{{\ttfamily arXiv:1203.6688
  [gr-qc]}}.

\bibitem{JuliansReview}
J.~Barbour, ``{Shape dynamics. An introduction},'' in {\em Quantum Field Theory
  and Gravity. Proc. Conference at Regensburg 2010}, F.~Finster, ed.
\newblock Birkh\"auser, 2012.
\newblock \href{http://arxiv.org/abs/1105.0183}{{\ttfamily arXiv:1105.0183}}.

%\bibitem{barbour:scale_inv_particles}
%J.~Barbour, ``{Scale-invariant gravity: particle dynamics},''
%  \href{http://dx.doi.org/10.1088/0264-9381/20/8/310}{{\em Class. Quant. Grav.}
%  {\bfseries 20} (2003) 1543--1570},
%  \href{http://arxiv.org/abs/gr-qc/0211021}{{\ttfamily arXiv:gr-qc/0211021}}.

\bibitem{BLMpaper}
J.~Barbour, M.~Lostaglio, and F.~Mercati, ``Scale anomaly as the origin of
  time,'' \href{http://arxiv.org/abs/1301.6173}{{\ttfamily arXiv:1301.6173
  [gr-qc]}}.


\bibitem{SaariBook}
D.~G. Saari, {\em Collisions, Rings, and other Newtonian N-body Problems}.
\newblock American Mathematical Society, Providence, RI, 2005.

\bibitem{LandauLifshitz1}
L.~D. Landau and E.~M. Lifshitz, {\em Course of theoretical physics}, vol.~1:
  Mechanics, ch.~II, sec. 10.
\newblock Pergamon Press, Oxford, 1976.

\bibitem{Niall_73}
N.~O'Murchadha and J.~James W.~York, Jr., ``Existence and uniqueness of solutions of
  the Hamiltonian constraint of general relativity on compact manifolds,''
  \href{http://dx.doi.org/10.1063/1.1666225}{{\em J. Math. Phys.} {\bfseries
  14} (1973) 1551--1557}.


\bibitem{DickeDimensions}
R.~Dicke, {\em Relativity, Groups, and Topology}.
\newblock Gordon and Breach, NY, 1964.

\bibitem{York:york_method_long}
J.~James W.~York, Jr., ``{Conformally invariant orthogonal decomposition of
  symmetric tensors on Riemannian manifolds and the initial value problem of
  general relativity},'' \href{http://dx.doi.org/10.1063/1.1666338}{{\em J.
  Math. Phys.} {\bfseries 14} (1973) 456--464}.


\bibitem{MatterPaper}
H.~Gomes and T.~Koslowski, ``Coupling shape dynamics to matter gives
  spacetime,'' \href{http://dx.doi.org/10.1007/s10714-012-1355-2}{{\em Gen.
  Rel. Grav.} {\bfseries 44} (10, 2012) 1539--1553},
  \href{http://arxiv.org/abs/1110.3837}{{\ttfamily arXiv:1110.3837 [gr-qc]}}.

\bibitem{Horava:lif_point}
P.~Ho{\v r}ava, ``{Quantum gravity at a Lifshitz point},''
  \href{http://dx.doi.org/10.1103/PhysRevD.79.084008}{{\em Phys. Rev.}
  {\bfseries D79} (2009) 084008},
  \href{http://arxiv.org/abs/0901.3775}{{\ttfamily arXiv:0901.3775 [hep-th]}}.

\bibitem{BarbourReductionistDoubts}
J.~Barbour, ``Reductionist doubts.'' Third-prize winning essay for the fqxi
  essay contest 'questioning the foundations', 2012.
\newblock \url{http://fqxi.org/community/essay/winners/2012.1#barbour}.

\bibitem{DiracHamiltonianDynamics}
P.~A.~M. Dirac, ``Generalized Hamiltonian dynamics,''
  \href{http://dx.doi.org/doi:10.1098/rspa.1958.0141}{{\em Proc. R. Soc. A}
  {\bfseries 246} (1958) 326--332}.

\end{thebibliography}
\end{document}